\def\real{{\tt I\kern-.2em{R}}} 
\def\nat{{\tt I\kern-.2em{N}}}     
\def\eps{\epsilon}
\def\realp#1{{\tt I\kern-.2em{R}}^#1}
\def\natp#1{{\tt I\kern-.2em{N}}^#1}
\def\hyper#1{\,^*\kern-.2em{#1}}
\def\monad#1{\mu (#1)}

\def\hyperrealp#1{{\tt ^*{I\kern-.2em{R}}}^#1} 

\def\hypernatp#1{{{^*{{\tt I\kern-.2em{N}}}}}^#1} 
\def\eskip{\hskip.25em\relax}

\def\Hyper#1{\hyper {\eskip #1}}
\def\leaderfill{\leaders\hbox to 1em{\hss.\hss}\hfill}
\def\srealp#1{{\rm I\kern-.2em{R}}^#1}

\def\pars{\par\smallskip}

\def\r#1{{\rm #1}}

\def\ref#1{$^{#1}$}

\def\m@th{\mathsurround=0pt}
\def\rightarrowfill{$\m@th \mathord- \mkern-6mu \cleaders\hbox{$\mkern-2mu 
\mathord- \mkern-2mu$}\hfil \mkern-6mu \mathord\rightarrow$}
\def\leftarrowfill{$\mathord\leftarrow
\mkern -6mu \m@th \mathord- \mkern-6mu \cleaders\hbox{$\mkern-2mu 
\mathord- \mkern-2mu$}\hfil $}
\def\noarrowfill{$\m@th \mathord- \mkern-6mu \cleaders\hbox{$\mkern-2mu 
\mathord- \mkern-2mu$}\hfil$}
\def\orgate{$\bigcirc \kern-.80em \lor$}
\def\andgate{$\bigcirc \kern-.80em \land$}
\def\inverter{$\bigcirc \kern-.80em \neg$}

\magnification=\magstep1
\tolerance 10000
\baselineskip  14pt
\hoffset=.25in
\hsize 6.00 true in
\vsize 8.75 true in
\centerline{\bf Modern Infinitesimal Analysis Applied to the Physical Metric dS and a}
\centerline{\bf Theoretical Verification of a Time-dilation Conjecture.}
\medskip 
\centerline{Robert A. Herrmann}\par
\centerline{9 FEB 2008}\par\bigskip
{\leftskip=0.5in \rightskip=0.5in \noindent {\it Abstract} In this paper, the modern theory of infinitesimals is applied to the General Relativity metric $dS$ and its geometric and physical meanings are rigorously investigated. Employing results obtained via the time-dependent Schr\"{o}dinger equation, gravitational time-dilation expressions are obtained and are shown to be caused by gravitationally altered photon interactions with atomic structures.\par}\par\medskip 
\noindent {\bf 1. Introduction.}\pars
\noindent In books that deal with General Relativity, one often sees the expression for the ``metric'' or line-element $dS\ (= ds).$ This symbolism usually appears on the left-hand side of an equation and the symbols $dx^j$ or $dx^\eta$, where $j,\ \eta$ vary from 1 to 4, appear on the right-hand side. So, the $dS$ is related to the set of $dx^j$. This, of course, does not indicate how this $dS$ should be interpreted in a physical sense. The meanings for these symbols were thought to be well understood when they were first displayed within mathematics. However, there was a certain level of confusion and even error in the physical interpretations that have developed since such interpretations were introduced. \par
Since the time of Newton through 1855, the entire period when basic ``calculus''  notions were introduced, the application of these symbols to geometry and physical science was based upon the ``intuitive'' behavior of what was conceived of as the ``infinitely small'' and how ordinary Euclidean geometry and physical measures are perceived. For example, in calculus textbooks written from 1700 - 1850s a space curve is defined as an ``infinite collection of infinitely small line segments.'' This is distinct from how we actually go about physically measuring the ``length'' of a curve segment. If you take a segment of a two dimensional curve, draw it on a piece of paper and lay a string on the entire path, then to measure the length you would lay out the string in a ``right line'' as they would call a ``straight line'' and place it on a ruler. Then simply read off the length of the curve in terms of a Euclidean unit.  \par

Students learned, by example, how to handle the mathematical symbols used to identify infinitely small measures - the ``little'' o. There were no exact rules for their behavior. However, in 1826, it was shown that the known rules used to argue in terms of the language of the infinitely small were mathematically inconsistent. This did not stop these inconsistent rules from being applied. The term infinitely small or the modern term infinitesimal is still used today as are the ``old'' rules for how such symbols are to be handled. For example, in [7] on page 134, McConnell writes, ``Let $P$ be a point whose coordinates are $x^r$ and let $Q$ be a neighboring point with coordinate $x^r +dx^r.$ If we denote the infinitesimal distance $PQ$ by 
$ds$, we call $ds$ the {\it element of length} etc.'' This can only mean that the $dx^r$ are infinitesimals, whatever they are. There is even a section entitled  ``Infinitesimal deformations,'' where he only states that these are ``{\it small}  homogenous deformations.'' He states that these deformations are ``infinitesimals of the first order.'' He neither gives a definition for the term ``infinitesimal'' or ``small'' nor any of the algebra for the ``small'' or the ``infinitesimals.'' The only way one could learn how to work with the {\it small} is to replicate his ``proofs'' and nothing more could be known about the notion of the {\it small.} 
Since this is not rigorous mathematics, it can only be hoped that what properties of the {\it small} that are gleaned from these and other statements made by those who write in differential geometry and physics as to how the {\it small} behaves will not lead to one of the known inconsistencies.\par

From these examples, it should be clear that to retain logical rigor within the calculus one needs to find the proper algebraic properties for the infinitesimals so that they can be properly manipulate and applied to the physical world. \pars

\noindent{\bf 2. The Modern Infinitesimals.}\pars

\noindent To rigorously understand $dS,$ a ``new'' set of numbers  - the actual {\it infinitesimals} - is adjoined to the real numbers $\real$ (or even to the complex numbers) and a new rigorous infinitesimal arithmetic is needed and the order $<$ needs to be properly extended. These new infinitesimal numbers are often denoted by lower case Greek letters and the collection of these new infinitesimals as adjoined to $\real$ is denoted by $\monad {0}.$ (Other new numbers are also adjoined to $\real$, and they are introduced as needed.) The basic arithmetic is not too difficult to comprehend. For this article, here are the necessary ``rules.'' \par

The ``addition'' and ``multiplication'' operations for the real numbers are extended to members of $\monad {0}$ as follows: Any nonempty \underbar{finite} sum or multiplication of members from $\mu(0)$ is a member of $\mu(0).$  Members of $\mu (0)$ satisfy all of the basic ``grouping'' rules for $\real$. For example, if $\alpha,\ \beta,\ \gamma \in \mu(0),$ then $\alpha(\beta + \gamma) = \alpha\beta + \alpha\gamma \in \mu (0).$ Let $a \not=0,$ $(-a,a)=\{x\mid (x \in \real)\land (-a < x < a)\},$ and let $f$ be any continuous function defined on $(-a,a)$ such that $f(0) =0$. Then applying the same ``function'' but extended to $\mu (0)$ will always yield members of $\mu(0).$ \par

Although it can be obtained from the abstract algebraic properties of $\mu(0)$, an important application of this continuous function idea is when $f(x) = gx,$ where $g \in \real$. Now substitute into this expression an infinitesimal $\eps$ and obtain $g\,\eps$ that is also a member of $\mu (0).$ {\it This fact is a very important requirement when members of $\mu(0)$ are used.} On the other hand, if $0\not=r\in \real$, then, as intuitively expected, $r + \eps\notin \mu(0)$, although it is a new number that behaves like a real number. The set $\mu (0)$ is shown to exist mathematically by various means, one of which is by pure abstract algebra as well as ideas from modern mathematical logic. Also, there is one important operation that is done with  members of $\mu (0)$ and gives interesting new numbers that are not in $\mu (0).$ Let $\eps\not= 0$ Then, although $1/\eps$ is one of these new numbers that is adjoined to $\real,$ $1/\eps$ is {\bf not} a member of $\mu (0).$ This is a significant fact. An additional significant fact is that, with respect to the extended order $<$, if $r$ is any real number and $\eps >0,$ then $r < 1/{\eps}.$ If $\eps < 0,$ then $1/\eps < r.$   \par

The major property for members of $\mu (0)$ is how the extended real number order behaves. 
Let $0< r\in \real$ and  $\eps \in \mu(0),$ then $\eps < r.$ If $r < 0,$ then $ r < \eps.$ Intuitively, the set $\mu (0)$ only contains the one real number $0$ and all other members of $\mu(0)$ ``crowd around'' $0$ ``closer than'' any other real number. If $f$ is a real valued function defined on all the real numbers $[a,b]=\{x\mid (x \in \real)\land(a\leq x \leq b)\}$ and there is a real number $B$ such that $\vert f(x)\vert  \leq B$ for each $ x \in [a,b]$, then, for any $\eps \in \mu(0)$ and any $x \in [a,b],\ \eps f(x) \in \mu(0).$ Any continuous $f$ defined on $[a,b]$ has this property. The behavior of the members of $\mu (0)$ appears to follow all of the notions associated with the ``small'' notion and the original definitions used in the calculus. Further, they should eliminate all of the known inconsistencies in the use of the terms ``infinitely small'' and ``infinitesimal'' throughout differential geometry and all other applications of the calculus. Also note that all ordinary real number arithmetic holds for the $\eps$ and $1/\eps$ numbers. \par 

Similar to Cantor's definition of the real numbers as equivalence classes of rational number Cauchy sequences, certain sequences of real numbers that converge to zero can be used to represent an infinitesimal. For example, consider the sequence with values $\{G_n= 1/n\},$ where $n$ is a nonzero member of natural numbers $\nat$ and where $G_0 = 0.$ Then the sequence $\{G_n\}$ can ``represent'' an infinitesimal. There is an equivalence class of such sequences that are related in a special way and the $G$ is one member.  Mathematically, there are actual formal mathematical objects that yield infinitesimals. But, what is the proper way to apply the calculus, using these now rigorously defined numbers, to real world applications?\pars\vfil\eject

\noindent{\bf 3. The Geometric and Physical Meaning of the dS.}\pars

\noindent({\bf In all that follows, all lower case Greek letters represent infinitesimals, where 0 is an infinitesimal.}) What is presented next could be classified as an overly-long explanation. However, if individuals want exposure to all of the correct steps that are usually hand-waved over in the ordinary course in calculus and physics, then what follows is necessary. To start, the most general form is not discussed, but the 4-dimensional ``form''  for a physical metric is considered. Further, the Einstein summation convention is not used. One form is 
$$(dS)^2 = dS^2= h_1g_1(dx^1)^2 + h_2g_2(dx^2)^2 + h_3g_3(dx^3) + h_4g_4(dx^4)^2,\eqno (1)$$
where the $x^1,\ x^2,\ x^3,\ x^4$ denote distinct variables that are used to denote a ``point'' name $(x^1,x^2,x^3,x^4)$ in spacetime and, when evaluated at point names, the functions  $g_j\geq 0,\ 1\leq j \leq 4.$ The $h_j = \pm 1.$ Note that the $dS^2$ is only consider, at the moment, as an abbreviation for this form. In this form, what do the $dx^i$ mean?\par

It is often stated that they represent a ``small'' (infinitesimal) change. But, physically or geometrically a change in what? Consider an interval of real numbers $[0,1].$ Take $u$ as a parameter that varies over $[0,1].$ Consider the four linear equations
$x^1 = a_1 + ub_1,\ x^2 = a_2 + ub_2, \ x^3 = a_3 + ub_3,\ x^4 = a_4 + ub_4.$ The collection of all four-tuples generated by these equations is termed as a ``linear path'' from the point with ``name'' $(a_1,a_2,a_3,a_4)$ to the point with name $(a_1+b_1,a_2+b_2,a_3+ b_3,a_4 +b_4).$ Suppose that $[0,1]$ contains all of the required real numbers as well as all of these new numbers; the non-negative infinitesimals (infinitely close but $\geq 0$) and all the ones that look like $r + \eps,$ where $0 \leq r + \eps\leq 1.$ The notation for this new view of $[0,1]$ is $\Hyper [0,1].$ \par

Let $\eps > 0.$ Consider a ``micro''-linear path from $(a_1,a_2,a_3,a_4)$ to $(a_1 + \eps b_1,a_2 + \eps b_2,a_3 + \eps b_3,a_4 + \eps b_4).$ These are the end-points of a micro-line segment obtained by varying  infinitesimal $\gamma$, where $0\leq \gamma\leq \eps,$ and letting $x^j = a_j + \gamma b_j,\ j =1,2,3,4.$  Consider the usual coordinate vector algebra and obtain the components $(\eps b_1,\eps b_2,\eps b_3,\eps b_4)$. Using the arithmetic of the infinitesimals, it follows that each of these components is an infinitesimal. \par

By definition, each $dx^j$ means, for these equations, differences of this type. Hence, $dx^j = \eps b_j$, where $j$ will always vary from 1 to 4, represents an infinitesimal (i.e. small) ``change'' in the point name. Now, letting each $g_j = 1$, then  
$$dS^2 = (\eps b_1)^2 + (\eps b_2)^2 +(\eps b_3)^2 +(\eps b_4)^2= \sum_{j =1}^4 (\eps b_j)^2 \eqno (2)$$ and using the arithmetic for these new numbers  
$$dS = \sqrt {(b_1)^2 + (b_2)^2 +(b_3)^2 +(b_4)^2}\,du,\eqno (3)$$ where $du$ is the positive infinitesimal $\eps$  and $dS$ is a non-negative infinitesimal.\par

{\leftskip=0.5in \rightskip=0.5in \noindent IF these point names are considered as names for points using a Cartesian coordinate system, which is not easily drawn unless coordinates are suppressed, and physically the equations represent a geometric linear path for such points, then this result might be classified as the infinitesimal Euclidean micro-path length for the micro-linear path from $(a_1,a_2,a_3,a_4)$ to $(a_1 + \eps b_1,a_2 + \eps b_2,a_3 + \eps b_3,a_4 + \eps b_4).$ This seems to be a ``transfer'' of the linear path length  notion from the non-infinitesimal to the infinitesimal world. \par}\par

The next step is to generalize this to non-linear paths that start at $u = 0$ and end at $u = 1.$  Suppose that $x^1 = f_1(u), x^2 = f_2(u), x^3= f_3(u), x^4 = f_4(u), \ u \in [0,1],$  where paths between different points can usually be written so that only the interval $[0,1]$ is employed. Can a collection of infinitesimal micro-linear paths be used and intuitively be joined together, so to speak, and the length of the entire path between the two points $(f_1(0),f_2(0),f_3(0),f_4(0)$ and $(f_1(1),f_2(1),f_3(1),f_4(1))$ be obtained? Maybe, but as  this demonstration progresses something else might be necessary. \par

Consider the definite integral applied to equations such as (3). As viewed from the infinitesimal world, all the usual definite integrals are independent from the infinitesimal used for the $du.$\par

{\leftskip=0.5in \rightskip=0.5in \noindent Intuitively, from the infinitesimal world, all ordinary integrals, relative to $[0,1],$ are but extensions of ordinary finite sums $\sum_{i =0}^n k(u_i)\Delta u$, where the integrand function $k$ is defined on $[0,1],$ has specific properties and the $u_i$ must be located in restricted positions. The major difficulty is in obtaining the required ``form'' $k(u_i)du,$ where $du \in \mu(0),$ and whether the integral yields the physical measure being considered.\par}\par  

The usual modeling approach requires the interval $[0,1]$ to be divided, in the infinitesimal world, into infinitesimally ``long'' pieces.  So as to conform to the usual notion, the symbol $du$ is used and is a positive infinitesimal.  There are ``numbers,'' $\Gamma,$ that behave in many ways like members of $\nat$ and if $x \in \real$, then $x < \Gamma$ and $1/\Gamma \in \mu(0).$  These objects satisfy the Newton and Leibniz notion of the ``infinitely large.'' Consider ${{1-0}\over {\Gamma}}= du$ as the distance between division points.  Consider all of the $\Hyper {[0,1]}$ division points 
$$\{0 = u_0,u_1=u_0 +du, u_2 = u_1 +du, u_3= u_2 +du,\ldots,u_\Gamma = u_{\Gamma-1} + du =1\},\ {\rm or}$$
$$\{0 = u_0, du = u_1, 2du = u_2, 3du = u_3,\ldots,(\Gamma -1)du, \Gamma du = 1\}. \eqno (4)$$
This gives a collection of (infinitesimal) $du$ long subintervals 
$$\{[0=u_0,u_1],[u_1,u_2], \ldots, [u_{\Gamma -1},u_\Gamma = 1]\}.\eqno (5).$$\par

For each of these subintervals, one has the corresponding micro-linear paths  
from 
\noindent $(k_1\sqrt {g_1(u_i)}f_1(u_i),k_2\sqrt {g_2(u_i)}f_2(u_i),k_3\sqrt {g_3(u_i)}f_3(u_i),k_4\sqrt {g_4(u_i)}f_4(u_i)$ to\hfil\break $(k_1\sqrt {g_1(u_i + du)}f_1(u_i +du)),k_2\sqrt {g_2(u_i + du)}f_2(u_i +du),k_3\sqrt {g_3(u_i + du)m}f_3(u_i +du),k_4\sqrt {g_4(u_i + du)}f_4(u_i +du)),$ where $i = 0,1,2,3,\ldots, \Gamma -1$ and $k_j = 1$ or $k_j= {\rm i} =\sqrt {-1}.$ The idea of using ``complex'' coordinate names and, hence, complex geometry is not new. It was introduced into differential geometry in 1822 [9]. Further, in relativity theory, it was introduced by Minkowski in about 1906. \par

Next, the method used to arrive at (2) is applied to each of these subintervals, where the primary square root is used. This gives 
$$dS_i = \sqrt {\sum_{j =1}^4 h_j(\sqrt {g_j(u_i +du)}f_j(u_i + du)- \sqrt {g_j(u_i)}f_j(u_i))^2}. \eqno (6).$$
Assume that $g_j$ and $f_j$ are each continuous on $[0,1].$ In this case, a major property for continuous functions is that each of terms in the right-hand side of (6) is an infinitesimal. Thus $dS_i\in \mu(0)$ for each $i=0,1,\ldots \Gamma -1.$
Notice that the $du$ does not appear outside the radical. However, intuitively, one might claim that the length of the original micro-line segments is being altered. \par

All of these $dS_i$ are added in order to obtain the form   
$$\sum_{i =0}^{\Gamma -1}dS_i = \sum_{i=0}^{\Gamma -1}\sqrt {\sum_{j =1}^4 h_j(\sqrt {g_j(u_i +du)}f_j(u_i + du)- \sqrt {g_j(u_i)}f_j(u_i))^2}. \eqno (7).$$
Once again if $h_jg_j(u) = 1,$ then (8) represents the Euclidean length of a micro-polygonal path. This is a collection of attached micro-line segments. In general, if some $0 \not= h_jg_j \not= 1,$ then this can be viewed as a shift to a different (possibly complex coordinate) polygonal path. (Such ``paths'' were first investigated in 1822 [9].) However, there are two difficulties. First, can the left-hand side of (7) be put into the proper form so that an integral can represent the values and does the $S$ actually measure the modified path length notion represented by the function $\ell(x,y)$?\par

The most basic property a definite integral displays is that it is ``additive.'' For a function like $\ell(x,y)$, this means that for three parameter intervals $[a,b],\ [b,c],\ [a,c],$ $\ell(a,b) + \ell(b,c) = \ell(a,c).$  If you assume this property, then there is another property that the  function $S$ must satisfy before it can represent 
$\ell$. This property requires that each infinitesimal $\ell_i(u_i, u_i +du)$ be ``closer" to its approximation $dS_i$ than just ``infinitely close.''  This concept states that for $S$   
and any non-zero infinitesimal $du$ there must exist an infinitesimal $\beta_i$ such that 
$$\ell(u_i, u_i + du) = (S(u_i + du) - S(u_i)) + \beta_i du. \eqno (8)$$\par

In general for other measures, the $S$ can be replaced with other appropriate functions $h$ such as $(h(u_i + du) - h(u_i))du.$ In this case, the statement is written in the form 
$${{\ell(u_i, u_i + du)}\over{du}} = {{(h(u_i + du) - h(u_i))du}\over{du}} + \beta_i = $$ 
$$(h(u_i + du) - h(u_i)) + \beta_i. \eqno (9)$$
Note that the wrong notation  for these statements is being employed. In all the above cases, each of the functions must be ``extended'' to the entire set $\Hyper {[0,1]}$ and a new symbol is used for these extended functions. This notation has not been used in order to minimize notation. If expression (8) holds, then $\ell$ and $S$ are said to be {\bf indistinguishable of order 1 or infinitely close of order 1.} [Note: If additional requirements are imposed  upon $\ell$ and $S$, then $\ell$ and $S$ automatically satisfy this requirement.] \par

Thus, it is assumed that $\ell$ and $S$ are infinitely close of order one and this is applied to the left-hand side (7). It is known that if $S$ can be obtained by an integral, then it is differentiable on $[0,1].$ Hence, assuming this, from the Fundamental Theorem of Differential Calculus in infinitesimal form there are $\alpha_i$ and $u_i' \in [u_i, u_{i+1}]$ such that $S(u_i + du) - S(u_i) = S'(u_i')du + \alpha_i du= dS_i +\alpha_idu.$ Under the assumption that $\ell$ and $S$ are infinitesimally close of order 1, then 
$$\ell(u_i,u_{i+1}) = S(u_i + du) - S(u_i) + \beta_i du = S'(u_i')du + (\alpha_i + \beta_i)du. \eqno (10)$$
$$\sum_{i=0}^{\Gamma -1}\ell(u_i,u_{i +1}) =\sum_{i=0}^{\Gamma -1}((S'(u'_i)du +(\alpha_i + \beta_i)du), \eqno (11)$$
where we note that $(\alpha_i + \beta_i) = \nu_i \in \mu(0).$ \par

Now comes the interesting fact about this type of ``summation.'' {\bf It behaves in many ways just like finite summation.} Since every nonempty finite set of real numbers contains a maximum number, the set of $\{\nu_i\}$ as $i$ varies from 
$0$ to $\Gamma -1$ contains a maximum $\nu$. In what follows, the triangle inequality holds for a *-sum from 0 to $\Gamma -1$ (the *- can be translated, in this case, as the word ``hyperfinite''). Further, $\vert \sum_0^{\Gamma-1}\nu\, du \vert = \vert \nu \vert \ \vert\sum_0^{\Gamma-1}du\vert =\vert \nu\vert (1)=\vert \nu\vert$ since there are $\Gamma$ constant $du$ terms in this series and $\Gamma \, du = 1.$ 
Thus, using the *-triangle inequality 
$$-\vert\nu\vert\leq \sum_{i=0}^{\Gamma -1}\ell(u_i,u_{i +1}) - \sum_{i=0}^{\Gamma -1}S'(u'_i)du \leq \vert \nu \vert, \eqno (12)$$
and, hence, $\sum_{i=0}^{\Gamma -1}\ell(u_i,u_{i +1})$ is infinitely close to $\sum_{i=0}^{\Gamma -1}(S'(u'_i)du.$ \par

The *-series  $\sum_{i=0}^{\Gamma -1}S'(u'_i)du=\sum_{i=0}^{\Gamma -1}dS_i$ is in the exact form needed to replace it with $\int_0^1 dS = \int_0^1 S'(u)du = S(1) - S(0)$ if $S'(u)$ is Riemann integrable. [The integral is actually infinitely close to such sums and to obtain $S(1) - S(0)$ the standard part operator is employed.] The function $S'$ is integable if it is continuous. This will be the case here. Hence, under this assumption $\int_0^1 S'(u)du = S(1)-S(0)$. But, $\sum_{i=0}^{\Gamma -1}\ell(u_i,u_{i +1}) = \ell(0,1)$ by additivity. Hence, $S(1)-S(0)$ and $\ell(0,1)$ are ``infinitely close'' as Newton might say. But they are real numbers and cannot be ``infinitely close'' unless they are equal. Hence, $\ell(0,1)= S(1)-S(0).$\par

For the right-hand side, additional properties will lead to the integral form. Consider the right-hand side of 
$$\sum_{i =0}^{\Gamma -1}dS_i=\sum_{i=0}^{\Gamma -1}\sqrt {\sum_{j =1}^4 h_j(\sqrt {g_j(u_i +du)}f_j(u_i + du)- \sqrt {g_j(u_i)}f_j(u_i))^2}. \eqno (13).$$
One of the properties for a continuous function defined on $[0,1]$ is that it is ``uniformally continuous.'' From the infinitesimal view point, this yields that for any
 $v_{ij} \in [u_i,u_{i+1}]$ there is an $\alpha_{ij}$ such that $\sqrt {g_j(v_{ij})} = \sqrt {g_j(u_i)} + \alpha_{ij}$ and a $\beta_{ij}$ such that $\sqrt {g_j(v_{ij})} = \sqrt {g_j(u_i +du)} + \beta_{ij}.$
Substituting and using the arithmetic of infinitesimal numbers yields
$$\sum_{i =0}^{\Gamma -1}dS_i=\sum_{i=0}^{\Gamma -1}\sqrt {\sum_{j =1}^4 h_jg_j(v_{ij})(f_j(u_i + du)- f_j(u_i)+ \gamma_{ij})^2} =$$ 
$$\sum_{i=0}^{\Gamma -1}\sqrt {\sum_{j =1}^4 h_jg_j(v_{ij})(f_j(u_i + du)- f_j(u_i)+ \nu_{ij}du)^2}.\eqno (14)$$\par

Assuming that each $f_j$ is continuously differentiable on $[0,1]$ (i.e. the path is smooth), apply the (extended) Mean Value Theorem for Derivatives. Thus, there is for each $du$ a $u_{ij}' \in [u_1,u_{i+1}]$ such that $f_j(u_i + du)- f_i(u_i) = f'_j(u_{ij}')du + \eps_{ij}du.$  Substituting yields that 
$$\sum_{i =0}^{\Gamma -1}dS_i = \sum_{i=0}^{\Gamma -1}\sqrt {\sum_{j =1}^4 h_jg_j(v_{ij})(f'_j(u_{ij}') + \eps_{ij}')^2}\,du. \eqno (15)$$
Squaring the terms inside the radical and using infinitesimal number arithmetic, (15) is re-written as
$$\sum_{i =0}^{\Gamma -1}dS_i = \sum_{i=0}^{\Gamma -1}\sqrt {\sum_{j =1}^4 h_jg_j(u_{ij}')(f'_j(u_{ij}'))^2+ \eps_{i}}\,du \eqno (16)$$
where each $v_{ij} =u_{ij}'$.\par
Taking a closer look at the expression under the radical, it follows that (16) can be re-written as 
$$\sum_{i =0}^{\Gamma -1}dS_i = \sum_{i=0}^{\Gamma -1}\sqrt {\sum_{j =1}^4 h_jg_j(u_{ij}')(f'_j(u_{ij}'))^2}\,du + \delta_idu, \eqno (17)$$ 
The expression under the radical is now in the proper form. Using the exact same method as used to obtain equation (12) yields 
$$-\vert\nu_1\vert \leq \left(\sum_{i =0}^{\Gamma -1}dS_i\right) -  \sum_{i=0}^{\Gamma -1}\sqrt {\sum_{j =1}^4 h_jg_j(u_{ij}')f'_j(u_{ij}')^2}\,du \leq \vert \nu_1\vert.\eqno (18)$$
This implies, since the integrals exist, that 
$$\ell(0,1) = \int_0^1 \sqrt {\sum_{j =1}^4 h_jg_j(u)f'_j(u)^2}du. \eqno (19)$$\par

Hence under the stated requirements, the modified path measure is what the metric is trying to display in differential form. Intuitively, the $\ell(0,1)$ is obtained by considering the $h_jg_j$ as modifying the micro-line segments for a *-polygonal path without a gravitational field present and uses standard Euclidean-styled length measures for this new curve. This is accomplished by simply shifting to a different *-polygonal path. This is equivalent, but not intuitively, to considering the length of each micro-line segment as being modified from the Euclidean length. In general, the calculus uses Euclidean-styled notions in the infinitesimal world to achieve behavioral results. But, here, there are a few problems in interpretation.  First, one of the terms is a ``time'' statement, and this needs to be interpreted as such in all cases within General Relativity. Then some of the $h_j$ may be $= -1.$ \pars 

\noindent{\bf 4. Two Basic General Relativity Interpretations.}\pars

\noindent Today, in physical science, it is often necessary that a specific method to measure quantities such a ``length'' and ``time'' be included in an interpretation. Einstein introduced special modes of measurement (i.e. Einstein or apparent time, and Einstein distances [2, p. 368, 417]) into his Special and General Theories of Relativity and the General Theory reduces, in a local infinitesimal sense, to the Special Theory. {\it Such measurements incorporate the basic properties for light propagation.} When this is done, the Special Theory ``chronotopic interval'' statement $dS^2 = c^2dt^2 - dr^2$ is obtained. This form is actually obtained by consider the behavior of ``light prorogation,'' where ``relative velocity'' is measured by a special method. The method is the ``radar method,'' the transmission of a light-pulse that is reflected from an object and returned to the sender. Such modes of measurement must be used to measure the $t$ and the $r$ in this expression. (The expressions for the Einstein measures are given in section 5.) Hence, in any of the General Relativity metrics this mode or an equivalent mode of measurement must be done. Although it was usually mentioned in the past [2, p. 368], today this fact seems never to be mentioned. Einstein time and distance use the constancy of the measured to-and-fro speed of light. Further, within the physical world, the wave property that this speed does not change relative to the speed of a source is used. Electromagnetic radiation paths, as modeled by photons, appear to be the only physical world entities that can be infinitesimalized and that yield easily observed effects. The light-propagation device that can be used for Einstein measurements within the infinitesimal world, at least as an analogue model for such physical behavior, is the ``infinitesimal light-clock'' [5]. The ``geometry'' used for Relativity is called {\it chronogeometry} [4].\par
 
It has been shown in [5], with respect to gravitational fields and relative velocities, that certain physical changes in behavior take place when behavior is compared to behavior where there are no gravitational or relative velocity effects. These changes are all produced by changes in the (analogue) infinitesimal light-clock measuring devices. This is further related to how simple behavior within a substratum (the ``medium'') yields behavior within our universe as modeled by infinitesimal light-clock behavior. It is interesting that the notion of a substratum absolute time and distance is still a valid notion and, indeed, the original Einstein derivation that yields the Special Theory uses such an absolute time notion. \par

From the viewpoint of infinitesimal analysis, a path within a gravitational field is equivalent to a *-polygonal (i.e. micro-linear) path within the substratum medium, that is being modified within the physical world by a gravitational field.  The micro-length of each micro-line segment is altered by a constant multiplication factor. {\it But, the same effect can be obtained by simply altering the *-polygonal micro-linear path itself. This yields the alteration in the micro-length via a Euclidean-styled measure.} The effect may not be any more mysterious than that if using complex coordinates is not considered as mysterious. \par

 However, General Relativity also applies to certain behavior that is not stated in terms of ``paths.'' In all of these cases, there are gravitational ``potentials'' involved. There are two approaches. One method is to simply drop the notion of a *-polygonal path and replace it with a ``*-sequence of infinitesimal *-linear effects'' (i.e. micro-linear effects). Note that the ``time'' coordinate is necessary due to how ``velocity'' or ``*-sequence of infinitesimal *-linear effects'' are measured or applied, respectively. This approach would apply to ``very small'' real time intervals. The second approach is to assume that the spatial point is fixed. This yields only the ``time'' term in the metric. The expression obtained is then related to other measures. \par

\noindent {\bf 5. ``Time'' Measurable Aspects of $dS.$}\pars
\noindent The numbers being expressed by (1) need to be associated with physical measures. As shown [5] and noted in [2, p. 368], the $t$ and $r$ must be Einstein (radar) measures using light prorogation properties. [Note: {\it In what follows, the ``times'' $t_1,\ t_2,\ t_3$ are all being viewed from the medium. Such ``times,'' that have been termed as ``epochs,'' should not be considered as physical world measures, as yet. In section 7, it is shown how all of these times are directly related to physical world time measurements. However, their meanings with respect light propagation still remain valid.}] The points $P$ and $R$ originally coincide where the ``clock'' values coincide. The expression ``clock'' means an infinitesimal light-clock or its counter values. \par

A light-pulse leaves position $P$ at ``clock'' time $t_1,$ arrives at position $R$ at $R$ ``clock'' time $t_2.$ A type of ``reflected'' light-pulse arrives back at $P$ at time $t_3.$ The Einstein time $t_E$ is obtained by considering the ``flight-time'' that results using the wave property that all to-and-fro standard measures for $c$ are not altered by the velocity of the source, {\it where the sources are considered at standard locations (i.e. not infinitely close).} This Einstein approach assumes that the light-pulse path-length from $P$ to $R$ equals that from $R$ back to $P.$ Thus, the Einstein flight-time used for the distance $r_E$ from $P$ to $R$ is $(t_3 - t_1)/2$. The Einstein time $t_E,$ used for the ``clock'' time $t_2$ at $R,$ satisfies $t_3 - t_E = t_E - t_1.$ The ``times'' $t_1,\ t_3$ are local time values observed within the infinitesimal part of the medium. This yields the Einstein measures $t_E$ and $r_E$ as follows:
$$t = t_E = {{t_3 + t_1}\over {2}},\ r_E = c{{t_3 -t_1}\over{2}}, .\eqno (20)$$ 
For comprehension, let $, x^1 = t.$ (Note that the $h_jg_j$ ``time'' position is often denoted as $g_4$.) Hence, consider (1) in the form 
$$dS^2= h_1g_1(dt_E^1)^2 + h_2g_2(dx_E^2)^2 + h_3g_3(dx_E^3)^2 + h_4g_4(dx_E^4)^2,\eqno (21)$$ 
where $dS^2 \geq 0.$\par

When the integral expression for $S$ is considered, under any ``regular'' coordinate transformation where the integral expression is correctly altered by the transformation, the path length is of a fixed numerical value relative to standard units. This does not hold for every conceivable coordinate transformation but only for those that carry the additional property of ``regularity.'' The formal regularity requirement is not examined in this article. \par

To investigate a further physical meaning for (21), the idea of a ``point particle'' is introduced. Such a particle may have physical measures attached to it, but it is a mere ``point location'' in spacetime as far as the particle coordinate name is concerned. Assume that there is a ``clock,''  attached to the particle. Further, the usual modeling method of assuming physically simple behavior within the medium is used. From the particle $P$, observations are made of another particle $Q$ at a different spacetime location as it ``moves'' relative to $P$ within spacetime. For these observations, particle $Q$ is assumed to ``move'' in the medium along a *-polygonal micro-linear path to the spacetime point $Q'.$ \par

For the beginning of each piece of the micro-line segment, the $P$ ``clock'' reads $t_1$ as the ``clock'' time for the ``to'' part of the to-and-fro light-pulse model for $Q.$ The $t_2$ is an unknown $Q$ ``clock'' reading and $t_3$ is the $P$ reading for the ``fro'' part. At the other end of the micro-line segment, the readings are $\overline{t_1},\ \overline{t_3}.$  If the functions $f_j$ that describe the spacetime location of $Q$ \underbar{relative to P} are defined in terms of $t_E,$ then the inverse function theorem implies that $\overline{t_E}  - t_E \in \mu(0).$ Notice that, for infinitesimal regions, simple dynamics is assumed and this would imply that $t_E = {t_2} + \beta_1,\ \overline{t_E} = \overline{t_2} + \beta_2$ for the micro-line segments.  Since this is all relative to the behavior of light-paths, then the proper (or local) Einstein time invariant for $P$ observations of the two spacetime positions $Q,\ Q',$ is the right-hand side of (19) divided by $c,$ where $u = t_E.$ But, what would this all become for two spacetime events occuring only to $P$ and $P'$ ?\par

Let $Q$ be located such that $t_3- t_1 =\eps$. Hence, $t_E =t_3 + \beta.$  This yields that for the Einstein times $dt_E = \overline{t_E} - t_E = \overline{t_3} - t_3 + \gamma= dt_3 +\gamma.$    
 The time $t_3$ is the time, as originally viewed from the medium. {\it It represents, where each $f_j$ is defined on standard $t \in [a,\overline{a}],$  the ``beginning'' for an event at $P$ composed of infinitesimal micro-linear changes occuring over a ``time'' interval.} (Some authors actually use, in this context, the phrase ``infinitesimal observers'' for observations at $P$.)\par 

The metric (21) needs to be expressed entirely in terms of $t$. To do this, let $x^1= cf_1(t),\ u = t$ and let $T$ be an additive function measuring the ``local elapsed time'' experienced by this specific type of "clock" located at $P$. As before, it is necessary to assume, in order to obtain the integral result, that the measure $T$ is infinitely close of order 1. Now using the previous infinitesimal arithmetic methods along with the continuity of $f_j'$ for the path functions $f_j$, where  $f_j(t_E) = f_j(t_3) + \alpha_j$ and $dt_E = dt_3 + \beta,$  the local elapsed time interval measured from the medium for the two  events $P,\ P'$ is, where $a=t_3,$  
$$(S(\overline{a}) - S(a))(1/c) = T(\overline{a})- T(a) = (1/c)\int_{a}^{\overline{a}}
 \sqrt {h_1g_1c^2f'_1(t)^2 +\sum_{j =2}^4 h_jg_j(t)f'_j(t)^2}dt. \eqno (22)$$
 \par

What is needed where the point aspects are not relative to paths? In such a case, there is no change in the spatial location. Hence, in this case, (22) reduces to the ``particle's medium view'' of the cumulated micro-linear \underbar{time} changes.    
$$T(\overline{\overline{a}})- T(a) = (1/c)\int_{a}^{\overline{a}}
 \sqrt {h_1g_1c^2f'_1(t)^2}dt =\int_{a}^{\overline{a}}
 \sqrt {h_1g_1f'_1(t)^2}dt,\eqno (23)$$
where $h_1g_1 \geq 0.$\par

The problem is that this is the medium view. What is needed is that (23) be related to other viewpoints. To do this, the medium view is compared with the medium view where the gravitational field reduces to the Special Theory that is ``infinitely close'' to the General Theory at a point in spacetime [2, p. 416]. This is but the Special Theory chronotopic interval expression. For an event that occurs at $P$, $f_1(t)= t,$ and note that $t_1 = t_3= t_E = t,$ from the medium viewpoint. Hence, (23) reduces to  
$$T(\overline{t})- T(t) = \int_{t}^{\overline{t}}
 \sqrt {h_1g_1}dt. \eqno (24)$$\par

From the chronotopic interval, letting $h_1 = 1$ and $dr =0,$ it follows that
$$dt^s = (1/c)dS = dT = \sqrt {g_1(t)}dt^m. \eqno (25)$$ 
(Letting $h_1 = 1$ is based upon the requirement that the spatial point is fixed and ``time'' is varying (a timelike metric) and that the metric reduces to the medium chronotopic interval metric and that there is no change in $h_1.$) A basic principle is that for infinitesimal regions the gravitational potentials are infinitely close to constants. The ``s'' means a medium (substratum)  ``clock'' located at $P,$ where gravity (or acceleration) affects the ``clock'' located at $P$ from the medium viewpoint. The chronotopic interval is used since locally the gravitational alterations produce measurements that satisfy this interval statement. The ``m'' indicates the coordinate ``clock'' measured time interval viewed from the medium where there are no gravitational field effects. In general, $g_1$ may be time dependent, but the spatial coordinate names are fixed. Obviously, $g_1$ is a unitless number. \par

Suppose that $g_1$ is not time depended (the field is ``static'') and, hence, it behaves like a constant, relative to ``time,'' at the point. Then this gives for $P$
$$\Delta t^s = \sqrt {g_1}\Delta t_P^m.\eqno (26)$$
\par
\noindent{\bf 6. Medium Time-dilation Effects.}\pars

\noindent How should equation (26) be interpreted? Consider another spatial point $R$ within the gravitational field, where for the two points $P,\ R$ the expression $g_1$ is written as $g_1(P),\ g_1(R),$ respectively. Considering the point effect at each pont and applying the relativity principle, this gives, in medium $t_3$ time, that
$${{\Delta t^s_P}\over{\sqrt {g_1(P)}}}=\Delta t^m = {{\Delta t^s_R}\over{\sqrt {g_1(R)}}} \eqno (27)$$
Equations like (27) are comparative statements. This means that identical laboratories are at $P$ and $R$ and they employ identical instrumentation, definitions, and methods that lead to the values of any physical constants. Since infinitesimal light-clocks are being used, standard ``clock'' values can take on any non-negative real number value. The $\Delta t^s_P,\ \Delta t^s_R$ represent the comparative view of the gravitationally affected ``clock'' behavior as observed from the medium where there are no gravitational effects. (The $1/\sqrt {g_1}$ removes the effects.) 
  Assume a case like the Schwarzschild metric where real $\sqrt {g_1} <1$. Consider two different locations $P, R$ along the radius from the ``center of mass.'' Then there is a constant $r_s$ such that 
$$\sqrt {1- {{r_s}\over{r_P}}}\ \Delta t_{R}^s\ ({\rm in\ R\!\!-\!digits}) = \sqrt {1- {{r_s}\over{r_R}}}\ \Delta t_{P}^s \ ({\rm in \ P\!\!-\!digits}). \eqno (28)$$
where $r_s \leq r_P,\ r_R.$
[The cosmological ``constant'' $\Lambda$ modification ($\Lambda$ is not assumed constant) is 
$$\sqrt {1 - {{r_s}\over{r_P}} - (1/3)\Lambda{{r_P^2}\over{c^2}}}\ \Delta t_{R}^s =\sqrt {1 - {{r_s}\over{r_R}} - (1/3)\Lambda_1{{r_R^2}\over{c^2}}}\ \Delta t_{P}^s.]\eqno (28)'$$\par

Of course, these (28) [(28)'] are comparisons that must be done with the same type of ``clocks.'' As an example for (28), suppose that $r_s/r_P = 0.99999$ and $r_R = 100,000r_P.$ Then $r_s/r_R = .000009999$. This gives 
$0.003162278 \Delta t_{R}^s = 0.999995 \Delta t_{P}^s$. Hence, $\Delta t_{R}^s = 316.2262 \Delta t_P^s.$ Thus, depending upon which ``change'' is known, this predicts that ``a change in the number of R-digits'' equals ``316.2262 times a change in the number of P-digits.'' Suppose that at $P$ undistorted information is received. Observations of both the ``P-clock'' and the ``R-clock'' digit changes are made. (The fact that it takes ``time'' for the information to be transmitted is not relevant since our interest is in how the digits on the ``clocks'' are changing.) Hence, if the ``clock'' at $P$ changes by 1-digit (the ``clock'' tick), then the change in the R-digits is 316.2262. The careful interpretation of such equations and how their ``units'' are related is an important aspect of such equations since (27) represents a transformation. Using a special `` clock'' property, if the ``R-clock'' changes its reading by 1, then at $P$ the ``P-clock'' shows that only 0.003162 ``P-clock'' time has passed. If you let $R = \infty,$ then $\Delta t_{R}^s = 316.2278 \Delta t_{P}^s$ and, in a change in the reading of 1 at $P$, the R-reading at $\infty$ is 316.2278.  Is this an incomprehensible mysterious results? No, since it is shown in [5] that the gravitational field is equivalent to a type of change in the infinitesimal light-clock itself that leads to this result. But, for our direct physical world, thus far, the answer is yes if there is no physical reason why our clocks would change in such a manner. The equation (28) [$(28)'$] must be related to physical clocks within the physical universe in which we dwell.\pars

\noindent{\bf 7. The Behavior of Physical Clocks.}\pars

\noindent Einstein did not accept general time-dilation for the gravitation redshift but conjectured that such behavior, like the gravitational redshift, is caused by changes within atomic structures rather than changes in photon behavior during propagation. This was empirically verified via atomic-clocks. To verify Einstein's conjecture theoretically and to locate the origin of this atomic-clock behavior, the comparative statement that $dt^s = \sqrt {g_1} dt^m$ is employed. Using special techniques and the time-dependent 
Schr\"{o}dinger equation, it is shown in [5] that certain significant energy changes within atomic structures are altered by gravitational potentials.  Once again, consider identical laboratories, with identical physical definitions, physical laws, construction methods etc. at two points $P$ and $R$ and within the medium. When devices such as atomic-clocks are used in an attempt to verify a statement such as (28), the observational methods to ``read'' the clocks are chosen in such a manner  that any known gravitational effects that might influence the observational methods and give method-altered readings is eliminated. For point $P$, let $E_P^s,$ denote measured energy. (The ``s'' always means gravitationally affected behavior and the ``m'' always means the medium view where there are no gravitational effects.) In all that follows, comparisons are made.  Using the principle of relativity, the following equation (29) (A) holds, in general,  and if $g_1$ is not time dependent, then (B) holds.\vfil\eject 
$${\rm (A)}\ \sqrt {g_1(P)}dE_P^s = dE^m= \sqrt {g_1(R)}dE_{R}^s , \ {\rm (B)}\ \sqrt {g_1(P)}\Delta E_P^s = $$ $$\Delta E^m =\sqrt {g_1(R)}\Delta E_{R}^s.\eqno (29)$$ 
This is certainly what one would intuitively expect. It is not strange behavior. Hence, in the case that $g_1$ is not time dependent, then   
$$\sqrt {g_1(P)}\Delta E_P^s = \sqrt {g_1(R)}\Delta E_{R}^s,\eqno (30)$$

Quantum mechanics states, at least for an atomic structure, that the total energy is controlled by the time-dependent Schr\"{o}dinger equation. For this application, equation (29) corresponds to the transition between energy levels relative to the ground state for the specific atoms used in atomic-clocks. But, {\it for this immediate approach, the atomic structures must closely approximate spatial points.} Further, at the moment that such radiation is emitted the electron is considered at rest in the medium and, hence, relative to both $P$ and $R$. The actual aspect of the time-dependent Schr\"{o}dinger equation that leads to this energy relation is not the spatial ``wave-function'' part of a solution, but rather is developed from the ``time-function part.''  \par

{\it The phrase ``measurably-local'' means, that for the measuring laboratory the gravitational potentials are considered as constants.}) Diving each side of (B) in (29) by Planck's (measurably-local) constant in terms of the appropriate units, yields for two observed spatial point locations $P, \ R$ that
 $$\sqrt {g_1(P)}\nu_P^s = \sqrt {g_1(R)}\nu^s_{R}, \eqno (31)$$
 Equation (31) is one of the expressions found in the literature  for the gravitational redshift [6, p. 154] but (31) is relative to medium ``clocks.'' Originally (31) was verified for the case where $\sqrt {g_1(R)} \ll \sqrt {g_1(P)}$ using a physical clock. Note that since the $P$ and $R$ laboratories are identical, then the numerical values for $\nu^s_P$ and $\nu^s_R$ as measured using the altered medium ``clocks'' and, under the measurably-local requirement, are identical. Moreover, (31) is an identity that is based upon photon behavior as ``clock'' measured. \par

What is necessary is that a comparison be made as to how equation (31) affects the measures take at $R$ compared to $P,$ or at $P$ compared to $R$.  Suppose that $\vert \nu^s_A \vert$ indicates the numerical value for $\nu^s_A$ at any point $A$.  To compare the alterations that occur at $P$ with those at $R$, $\vert \nu^s_R \vert_P$ is symbolically substituted for the $\nu^s_P$ and the expression $\sqrt {g_1(P)}\vert \nu^s_R \vert_P = \sqrt {g_1(R)}\nu^s_{R}$ now determines the frequency alterations expressed in $R$  ``clock'' units.  As will be shown for specific devices, this is a real effect not just some type of illusion. {\it This substitution method is the general method used for the forthcoming ``general rate of change'' equation.} As an example, suppose that for the Schwarzschild metric $R = \infty$ and let $\vert \nu^s_R \vert = \nu_0.$ Then $\nu^s_\infty = \nu = \sqrt {1 - r_s/r_P}\,\nu_0.$ This result is the exact one that appears in [1, p. 222]. 
However, these results are all in terms of the behavior of the ``clocks'' and how their behavior ``forces'' a corresponding alteration in physical world behavior and not the clocks used in our physical world. These results need to be related to physical clocks. \par

Consider atomic-clocks. At $P$, the unit of time used is related to an emission frequency $f$ of a specific atom. Note that one atomic-clock can be on the first floor of an office building and the second clock on the second-floor or even closer than that. Suppose that the identically constructed atomic-clocks use the emission frequency $f$ and the same decimal approximations are used for all measures and $f$ satisfies the measurably-local requirement. The notion of the ``cycle'' is equivalent to ``one complete rotation.'' For point-like particles, the rotational effects are not equivalent to gravitational effects [8, p. 419] and, hence, gravitational potentials do not alter the ``cycle'' unit C. Using the notation ``sec.'' to indicate a defined atomic-clock second of time, the behavior of the $f$ frequency relative to the ``clocks'' requires, using equation (31), that 
$$\sqrt {g_1(P)}{{1\r C}\over{{\rm P\!\!-\!sec.}}}=\sqrt {g_1(R)}{{1\r C}\over{{\rm R\!\!-\!sec.}}}.\eqno (32)$$
$$\sqrt {g_1(P)}{{1}\over{{\rm P\!\!-\!sec.}}}= \sqrt {g_1(R)}{{1}\over{{\rm R\!\!-\!sec.}}}.\eqno (32)'$$
For measurably-local behavior, this unit relation yields that  
$$\sqrt {g_1(P)}(\overline{t_R} - t_R)({\rm R\!\!-\!sec.})= \sqrt {g_1(R)}(\overline{t_P} - t_P)({\rm P\!\!-\!sec.}).\eqno (33)$$
Hence, in terms of the atomic-clock seconds of measure 
$$\sqrt {g_1(P)}\Delta t_R= \sqrt {g_1(R)}\Delta t_P.\eqno (34)$$\par
Equation (34) is identical with (28), for the specific $g_1$, and yields a needed correspondence between the ``clock'' measures and the atomic-clock unit of time.   Corresponding ``small'' atomic structures to spatial points, if the gravitational field is not static, then, assuming that the clocks decimal notion is but a consistent approximation,  (34) is replaced by a (28) styled expression 
$$ \sqrt {g_1(P,t_P)}dt^s_R = \sqrt {g_1(R,t_{R})}dt^s_{P}\eqno (35)$$
and when solved for a specific interval correlates directly to atomic-clock measurements. Also, the Mean Value Theorem for Integrals yields 
$$\sqrt {g_1(R,t'_P)}(\overline{t_R} - t_R) = \sqrt {g_1(R,t'_{R})}(\overline{t_{P}} - t_{P}), \eqno (36)$$ 
for some $t'_P \in [t_P,\overline{t_P}],\ t'_{R} \in [t_{R},\overline{t_{R}}].$ 
Equations (34), (35) and (36) replicate, via atomic-clock behavior, the exact ``clock'' variations obtained using the medium time, but they do this by requiring, relative to the medium, an actual alteration in physical world photon behavior. The major interpretative confusion for such equations is that the ``time unit,'' as defined by a specific machine, needs to be considered in order for them to have any true meaning. As mentioned, the ``unit'' notion is often couched in terms of ``clock or observer'' language. The section 6 illustration now applies to the actual atomic-clocks used at each location. 
\par

For quantum physically behavior, how any such alteration in photon behavior is  possible depends upon which theory for electron behavior one choices and some accepted process(es) by which gravitational fields interaction with photons. Are the alterations discrete or continuous in character? From a quantum gravity viewpoint, within the physical world, they would be discrete if one accepts that viewpoint. This theoretically establishes the view that such changes are real and are due to ``the spacings of energy levels, both atomic and nuclear, [that] will be different proportionally to their total energy'' [3, pp. 163-164]. Further, ``[W]e can rule out the possibility of a simple frequency loss during propagation of the light wave. . . .'' [8, p. 184]. This  gravitational photon frequency redshift is not the only redshift that occurs in the behavior of electromagnetic radiation. For example, for the behavior of photons, there is the derivable Special Theory alterations as well as the cosmic ``redshift.''\par

Although the time-dependent Schr\"{o}dinger equation applies to macroscopic and large scale structures via the de Broglie ``guiding-wave'' notion, the equation has not been directly applied, in this same manner, to such structures since they are not spatial points. However, it does apply to all such point-approximating atomic structures since it is the total energy that is being altered. One might conclude that for macroscopic and large scale structures there would be a cumulative effect for a collection of point locations. Clearly, depending upon the objects structure, the total effect for such objects, under this assumption, might differ somewhat at different spatial points.  However, the above derivation that leads to (34) is for the emission of a photon ``from'' an electron and to simply extend this result to all other clock mechanisms would be an example of the model theoretic error in generalization unless some physical reason leads to this conclusion.\par

Equation (34) is based upon emission of photons. Throughout all of the atomic and subatomic physical world the use of photon behavior is a major requirement in predicting physical behavior, where the behavior is not simply emission of the type used above. This tends to give more credence to accepting that, under the measurably-local requirement, each material time rate of change, where a physically defined unit U that measures a $Q$ quality has not been affected by the gravitational field, satisfies 
$$\sqrt {g_1(P)}\Delta Q_P {\rm\ in\ a \ P\!\!-\!sec.} = \sqrt {g_1(R)}\Delta Q_R\ {\rm\ in\ an \ R\!\!-\!sec.},$$
 $$\Delta Q_R\ {\rm\ in\ an \ R\!\!-\!sec.}={{\sqrt {g_1(P)}}\over{\sqrt {g_1(R)}}}\vert \Delta Q_R\vert_P, \eqno (37)$$
Equations (37) give a comparative statement as to how gravity alters such atomic-clock time rates of change including rates for other types of clocks.\par

 Prior to 1900, it was assumed that a time unit could be defined by machines that are not altered by the earth's gravitational field. However, this is now known not to be fact and as shown in [5], such alterations in machine behavior is  probably due to an alteration in photon behavior associated with a substratum stationary source that undergoes two types of physical motion, uniform or accelerative. There is a non-reversible substratum process that occurs and that alters photon behavior as it relates to the physical world. These alterations in how photons physically interact with atomic structures and gravitational fields is modeled (mimicked) by the defining machines that represent the physical unit of time, when the mathematical expressions are interpreted. The observed accelerative and relative velocity behavior is a direct consequence of this non-reversible process. As viewed from the substratum, every uniform velocity obtained from the stationary first requires acceleration. This is why the General Theory and the Special Theory are infinitesimally close at a standard point. \par
It is claimed by some authors that regular coordinate transformations for the Schwarzschild solution do not represent a new gravitational field but rather allows one to investigate other properties of the same field using different modes of observation. When such transformations are discussed in the literature another type of interpretation appears necessary [5, p. 155-159]. Indeed, what occurs is that the original Schwarzschild solution is rejected based upon additional physical hypotheses for our specific universe that are adjoined to the General Theory. For example, it is required that certain regions not contain physical singularities under the hypotheses that physical particles can only appear or disappear at chosen physical ``singularities.'' Indeed, if these transformations simply lead to a more refined view of an actual gravitational field, then the conclusions could not be rejected. They would need to represent actual behavior. One author, at least, specifically states this relative to the Kruskal-Szekeres transformation. In [10, p. 164], Rindler rejects the refined behavior conclusions that would need to actually occur within ``nature.''  ``Kruskal space would have to be {\it created in toto}: . . . . There is no evidence that full Kruskal spaces exist in nature.'' \par
One way to interpret the coordinate transformation that allows for a description of ``refined'' behavior is to assume that such described behavior is but a ``possibility'' for a specific gravitational field and that such behavior need not actually occur. This is what Rindler appears to be stating. But, since such properly applied coordinate transformations also satisfy the Einstein-Hilbert gravitational field equations, then, from the medium view, using collections of such ``possibilities'' is equivalent to considering different gravitational fields. For the medium view of time-dilation, this leads to different alterations in the atomic-clocks for each of these ``possibilities.'' \par 

These results, as generalized to the behavior exhibited by appropriate physical devices, imply that no measures using these devices can directly determine the existence of the medium. Although Newton believed that infinitesimal values did apply to ``real'' entities and, hence, such measures exist without direct evidence, there is a vast amount of indirect evidence for existence of such a medium.\pars

\centerline{\bf References}
\noindent [1] Bergmann, G., Introduction to the Theory of Relativity, Dover, New York, 1976.\par
\noindent [2] Craig, H. V., Vector and Tensor Analysis, McGraw-Hill, New York, 1943.\par
\noindent [3] Cranshaw, T. E., J. P. Schiffer and P. A. Egelstaff, Measurement of the red shift using the M\"{o}ssbauer effect in $\rm Fe^{57},$ Phys. Rev. Letters 4(4)(1960):163-164,\par
\noindent [4] Fokker, Albert Einstein, inventor of chronogeometry, Synth\'{e} se 9:442-444.\par
\noindent [5] Herrmann, R. A., Nonstandard Analysis Applied to Special and General Relativity - The Theory of Infinitesimal Light-Clocks, 1992,93,94,95.\hfil\break http://arxiv/abs/math/0312189 \par
\noindent [6] Lawden, D. F., An Introduction to Tensor Calculus, Relativity and Cosmology, John Wiley \& Sons, New York, 1982.\par
\noindent [7] McConnell, A. J., Applications of Tensor Analysis, Dover, New York, 1957.\par
\noindent [8] Ohanian, H. and R. Ruffini, Gravitation and Spacetime, W. W. Nortin Co., New York, 1994.\par
\noindent [9] Poncelet, V., Trait\'{e} des propri\'{e}t\'{e}s projective des figures, 1822.\par
\noindent [10] Rindler, W., Essential Relativity, Springer-Verlag, New York, 1977. \par
\bigskip
\noindent USNA 43017 Daventry Sq, South Riding, VA 20152.

\end